\documentclass[reprint, superscriptaddress, twocolumn, amsmath, amssymb, prl]{revtex4-2}

\usepackage{graphicx}
\usepackage{bm}
\usepackage[colorlinks=true, citecolor=magenta, linkcolor=magenta]{hyperref}

\usepackage{mathrsfs}
\usepackage{color}
\usepackage{soul}
\usepackage{braket}

\usepackage{lipsum}
\usepackage{xargs}
\usepackage[colorinlistoftodos,prependcaption,obeyDraft]{todonotes}
\usepackage{setspace}

\usepackage{float}
\usepackage{booktabs}

\usepackage{flushend}

\newcommand{\Op}[1]{\ensuremath{\hat{#1}}}

\newcommand{\Abs}[1]{\vert#1\vert}

\newcommand{\Avg}[1]{\ensuremath{\langle #1 \rangle}}

\begin{document}

\title{Dicke State Generation and Extreme Spin Squeezing via Rapid Adiabatic Passage}

\def\ARL{DEVCOM Army Research Laboratory, Adelphi, MD 20783}
\def\MIT{Department of Physics, MIT-Harvard Center for Ultracold Atoms, and Research Laboratory of Electronics, Massachusetts Institute of Technology, Cambridge, MA 02139}
\def\ULM{Institute of Quantum Physics, University of Ulm, Ulm, Germany}
\def\Stevens{Department of Physics, Stevens Institute of Technology, Hoboken, NJ 07030}

\author{Sebastian C. Carrasco}
  \email{seba.carrasco.m@gmail.com}
  \affiliation{\ARL}

\author{Michael H. Goerz}
  \affiliation{\ARL}

\author{Svetlana A. Malinovskaya}
  \affiliation{\Stevens}

\author{Vladan Vuleti\'c}
  \affiliation{\MIT}

\author{Wolfgang Schleich}
 \affiliation{\ULM}

\author{Vladimir S. Malinovsky}
  \affiliation{\ARL}

\date{\today}

\begin{abstract}
Considering the unique energy level structure of the one-axis twisting Hamiltonian in combination with standard rotations, we propose the implementation of a rapid adiabatic passage scheme on the Dicke state basis. The method permits to drive Dicke states of the many-atom system into entangled states with maximum quantum Fisher information. The designed states allow to overcome the classical limit of phase sensitivity in quantum metrology and sensing. We show how to generate superpositions of Dicke states, which maximize metrological gain for a Ramsey interferometric measurement. The proposed scheme is remarkably robust to variations of the driving field and has favorable time scaling, especially for small to moderate ($\sim 1000$) number of atoms, where the total time does not depend on the number of atoms.
\end{abstract}

\maketitle

Quantum sensors have the potential to surpass their classical counterparts~\cite{TakanoPRL2009, SewellPRL2012, MuesselPRL2014, cox2016deterministic, HostenN2016, Pedrozo-penafielN2020, ColomboNP2022} and reach the fundamental quantum precision limit, the Heisenberg limit (HL)~\cite{PezzePRL2009, ZwierzPRL2010}, by fully exploiting non-classical properties of matter. In that limit, measurement precision scales proportional to the number of atoms. In contrast, the standard quantum limit (SQL) scales proportionally to the square root of the number of atoms. To achieve enhanced scaling, it is imperative to find robust ways to create ultra-sensitive entangled quantum states and engineering protocols to utilize them. The quantum advantage thus obtained will boost the precision of interferometric devices such as accelerometers~\cite{RaithelQST2022}, gyroscopes~\cite{GustavsonCQG2000, JarmolaSA2021}, and gravimeters~\cite{PetersN1999}. Further applications include the search for dark matter~\cite{derevianko2014hunting}, timekeeping~\cite{GuinotM2005}, gravitational wave detection~\cite{kolkowitz2016gravitational}, geodesy~\cite{mehlstaubler2018atomic, grotti2018geodesy, takamoto2020test}, and tests of the fundamental laws of physics~\cite{safronova2018search, safronova2019search} -- all fields where ultraprecise metrology plays the crucial role.

One common method for creating collective entanglement is through an effective one-axis twisting (OAT) Hamiltonian~\cite{kitagawa1993squeezed}, often engineered by exploiting the nonlinear interaction between the atoms and the light inside a cavity~\cite{schleier2010squeezing, leroux2012unitary, braverman2019near, ColomboNP2022, LiPRXQ2022}. That Hamiltonian squeezes the quantum state quasiprobability distribution, creating non-classical correlations that reduce the variance of one measurement quadrature while increasing the variance in the orthogonal direction. Thus, squeezing can enhance precision in Ramsey interferometric measurements~\cite{MaPR2011}. The maximum metrological gain in the context of Ramsey interferometry is achieved with particular squeezed states known as extreme spin squeezed (ESS) states~\cite{bloch2004control, toth2014quantum, sorensen2001entanglement, CarrascoPRA2022}. 
We consider the dynamics of the system in the Dicke basis, the set of eigenstates $\ket{S, m}$ of the operator $\Op{S}_z$.
In general, ESS states are a superposition of Dicke states, but as the squeezing increases, they gravitate to a single Dicke state.

Collective rotations generate transitions between Dicke states, which in combination with applications of the OAT Hamiltonian steer the $N$-atom system towards the desired ESS states. One widely used control scheme is to use a fixed-area resonant pulse scheme, that is, a train of Rabi pulses~\cite{ChiowPRL2011, KovachyN2015}. Precise control of the pulse power and duration is required for this method to be effective, since errors accumulate with the pulse train~\cite{Goerz2023Robust}.

An alternative is to use rapid adiabatic passage (RAP) to generate the state transitions~\cite{malinovsky2003momentum, UnanyanPRL2001, PeikPRA1997,KuznetsovaPhysScr2014,MalinovskayaOptLett2017}. In this case, the transition frequency sweeps through the resonance with the excitation frequency as in the Landau-Zener model~\cite{RubbmarkPRA1981}, and the frequency chirp leads to a robust population transfer~\cite{VitanovPRA1999,PachniakSRep2021}. Shortcuts to adiabaticity may be used to speed up the process~\cite{ChenPRL2010, Guéry-OdelinRMP2019}.

Here, we propose an implementation of the RAP method to create extreme spin-squeezed states and pure Dicke states. We consider $N$ non-interacting two-level atoms or spin one-half particles under the action of the Hamiltonian
\begin{equation} \label{Hamiltonian}
    \Op{H} = \chi \Op{S}_z^2 +  \beta (t)  \Op{S}_z + \Omega (t) \Op{S}_x\,,
\end{equation}
where $\Op{S}_j$ are the components of the collective spin operator, $j= x, y, z$. The first term is the entangling OAT interaction, and $\chi$ is the shearing parameter. The second and third terms are rotations around the $z$ and $x$ axes. We are addressing the situation in the limit where dissipation is weak, and the system is well represented by the Hamiltonian of Eq.~\eqref{Hamiltonian}. In the context of the Hamiltonian implementation using the interaction between atoms and light in a cavity, this case corresponds to the limit of very strong cavity coupling. The Hamiltonian can also be realized in superconducting qubits~\cite{PezzèRMP2018} or exploiting the Ising interactions between trapped ions~\cite{franke2023quantum} and Rydberg atoms~\cite{eckner2023realizing, bornet2023scalable}.

The equation of motion for the probability amplitude being in a Dicke state $\ket{S, m}$ is
\begin{equation} \label{motion}
i \dot{a}_m = E_m(t)  a_m + \frac{\Omega (t)}{2} \left(\zeta_+ a_{m+1} + \zeta_- a_{m-1}\right)\,,
\end{equation}
where $E_m = \chi m^2 +  \beta (t)  m$ is the state energy, $\zeta_{\pm} = [(S\mp m) (S\pm m+~1)]^{1/2}$ are transition elements, and $m=-S, -S+1, \ldots, S-1, S$ for $N=2S$ particles. We focus here on an even number $N$ of atoms, for which there is a unique ground state $\ket{S, 0}$~\footnote{For odd $N$, the two lowest Dicke states $\ket{S,\pm1/2}$ are degenerate, which affects the ESS states. Although the fundamental technique for even and odd $N$ is the same, there are subtle differences that will be analyzed in future work.}.

According to Eq.~\eqref{motion}, only neighboring Dicke states are coupled. Therefore, it is possible to generate successive transitions $m = n \rightarrow n \pm 1$,  or to create a superposition of several Dicke states by properly choosing the time-dependent function $\beta(t)$ in Eq.~\eqref{Hamiltonian} and controlling the duration of the field $\Omega (t)$. For example, we can start from the coherent spin state (CSS) $\ket{S, S}$ and then use various control methods~\cite{Sola_AAMO2018} to prepare a desired Dicke state or other correlated quantum states.

To evaluate the usefulness of a state for high-precision measurement, we use the quantum Fisher information (QFI)~\cite{BraunsteinPRL1994,EscherNatPhys2011,TothJPhysA2014}. For pure states, the QFI of a state $\ket{\psi}$ is $\mathcal{F}_j =4 (\braket{\psi|\Op{S}_j^2|\psi} - \braket{\psi|\Op{S}_j|\psi}^2)$, with $j=x,y,z$. For the Dicke state $\ket{S, m}$, we find $\mathcal{F}_z = 0$ and $\mathcal{F}_{x,y} = 4 \Delta S_{x,y}^2 = N^2/2 - 2 m^2 + N$. Thus, Dicke states are not sensitive to perturbations proportional to $S_z$ (due to their axial symmetry on the Bloch sphere). In contrast, $x,y$-components depend on $m^2$ and demonstrate a scaling transition from the SQL, $\mathcal{F}_{x,y}  = N$ for the $\ket{N/2, N/2}$ state (a CSS), to  $\mathcal{F}_{x,y}  = N^2/2 + N$ for the $\ket{N/2, 0}$ Dicke state. According to the Cramer-Rao bound, the maximum precision of a phase estimation is bounded by the QFI as  $\Delta \varphi^2_{x,y,z} \geq 1/\mathcal{F}_{x,y,z}$~\cite{BraunsteinPRL1994}. Thus $\ket{N/2, 0}$ is reaching the HL scaling for $x$ and $y$ (up to the pre-factor $1/2$) for $N \gg 1$.

To prepare the $\ket{N/2, 0}$ state via RAP, we apply the linear-chirping function $\beta (t)=\alpha t u(-t) $, i.e., the chirp rate $\alpha$ stops at $t=0$ due to the Heaviside step function $u(-t)$. In this case, the linear chirping tunes the transitions between the adjacent Dicke state to the resonance, and appropriate $\Omega(t)$ efficiently transfers population from the initial CSS to a target Dicke state or Dicke state superposition.

\begin{figure}
  \centering
  \includegraphics{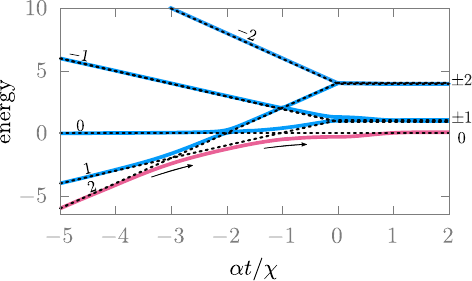}
  \caption{\label{fig:crossings} Example of the energy levels of the five lowest Dicke states as a function of time. The solid lines represent the instantaneous eigenvalues of the Hamiltonian in Eq.~\eqref{Hamiltonian} (adiabatic picture), and the dashed curves are the diagonal values, $E_m$, (diabatic picture). The coupling pulse goes up to $\Omega_{\text{max}} = 0.4 \chi$. It starts to turn off at $t=0$ and is entirely off at $\alpha t / \chi=1$. The chirp rate is $\alpha = 0.1 \chi^2$. }
\end{figure}

Figure~\ref{fig:crossings} presents
both the diabatic and adiabatic pictures of the multiple sequential crossings between the state energies, $E_m(t)$, which become avoided crossings due to the coupling $\Omega (t)$. The crossing time between adjacent Dicke states $m$ and $m-1$ is $t_{m,m-1}=\chi(1-2m)/\alpha$, providing resonances between adjacent Dicke states with period $\tau= 2 \chi/\alpha$. This comes from the interplay of the quadratic and linear structure of $\Op{S}_z^2$ and $\Op{S}_z$ eigenvalues, in complete analogy with the RAP between momentum states using frequency-chirped standing-wave fields~\cite{malinovsky2003momentum,Goerz2023Robust}.

In the adiabatic limit, each sequential avoided crossing can be considered independently, and the Dicke state population dynamics is described by the well-known Landau-Zener model~\cite{RubbmarkPRA1981}. Therefore, if at $t=-\infty$ the whole $N$-atom system population is in the CSS (single Dicke state  $\ket{N/2, N/2}$), then, according to the adiabatic theorem, the total evolution of the system happens in the single adiabatic state (the lowest solid line in Fig.~\ref{fig:crossings}). Since the chirp stops at $t=0$ and the coupling $\Omega (t)$ is turned off soon after, the last avoided crossing is between the Dicke states $\ket{N/2, 1}$  and $\ket{N/2, 0}$. Therefore, when adiabatic conditions are satisfied for all sequential crossings, the system population will be efficiently transferred to the target Dicke state $\ket{N/2, 0}$ at the final time. This qualitative picture is independent of the number of atoms, as long as adiabaticity is maintained ($\Omega^2_{\text{max}}/\alpha \gg 1$).

\begin{figure}
  \centering
  \includegraphics[width=\columnwidth]{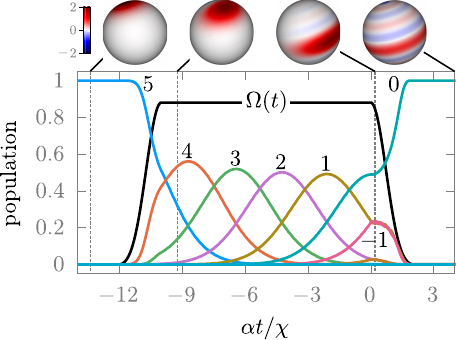}
  \caption{\label{fig:to_m=0} Population dynamics of the ten-atom Dicke states. The solid black line shows the coupling pulse shape, $\Omega (t)$. The chirp rate is $\alpha = 0.1 \chi^2$.}
\end{figure}
In Fig.~\ref{fig:to_m=0}, we show the dynamics of the Dicke state populations using RAP from $\ket{5, 5}$ to $\ket{5, 0}$ with fidelity $\epsilon^2 = 0.9996$. To be realistic, we are turning $\Omega(t)$ on and off with a Blackman shape. The value at the plateau is $\Omega_{\text{max}} = 0.88 \chi$. There is a residual population of the $\ket{5, \pm1}$  states building up at the end, which returns to the target state $\ket{5, 0}$ as the pulse $\Omega(t)$ is turned off. This transient effect is known as ``adiabatic population return for off-resonant excitation schemes''~\cite{Sola_AAMO2018}.

The results shown in Fig.~\ref{fig:to_m=0} are extremely robust to variations in the chirp rate $\alpha$ and the coupling strength $\Omega(t)$, see the Supplementary Material. Indeed, it is sufficient to have a slow turn-on before the first crossing, between energies $E_5(t)$ and $E_4(t)$ of the $\ket{5, 5}$ and $\ket{5, 4}$ Dicke states here. We set the plateau to start at $t_1=-N \chi/\alpha$ with a switch-on time of $t_\text{on} = 2 \chi/\alpha$, and choose $t_2=0$ as the plateau end with $t_\text{off} = 2 \chi/\alpha$ for the switch-off time. A time delay of the plateau end to $t_2=\chi/\alpha$ slightly reduces the fidelity to $\epsilon^2 = 0.9992$.  Increasing the number of atoms $N$ requires only an earlier start of the plateau time by a corresponding number of periods $\tau$ to accommodate more Dicke state crossings. The adiabatic picture in Fig.~\ref{fig:crossings} is valid for an arbitrary number of the atoms for the target Dicke state $\ket{N/2,0}$. It is also possible to choose any other Dicke state as a target, which can be efficiently prepared with high fidelity by applying the same excitation scheme. To selectively prepare another Dicke state $\ket{N/2,n}$, we need to adjust the plateau duration time so that the last avoided crossing is between states $\ket{N/2,n+1}$ and $\ket{N/2,n}$.

At the top of Fig.~\ref{fig:to_m=0}, we show a Wigner representation~\cite{DowlingPRA1994, KoczorPRA2020} \textcolor{red}{(see Supplementary Material)} of the system state on the generalized Bloch sphere at selected times. There are fringes indicating atomic coherence. Note that there is a reduced variance in the $z$-direction at the final time. In fact, the variance $\Delta \Op{S}_z^2$ is zero for any Dicke state, including the $\ket{N/2,0}$ state. However, there is no clear orientation of the total spin, the state is symmetric in the $x-y$ plane, and the mean spin components $\braket{\Op{S}_{x,y,z}}$ of the state are zero. Therefore, standard Ramsey interferometry with this state has zero contrast. However, it is possible to design another measurement scheme that can utilize the full quantum advantage of the Dicke state $\ket{N/2,0}$ {using a twist-and-turn strategy to decode the phase imprinted in the quantum state after free evolution} as in 
~\cite{KaubrueggerPRX2021,Kaubruegger2023}.
\begin{figure}
  \centering
  \includegraphics[width=\columnwidth]{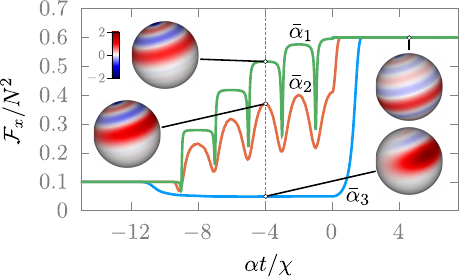}
  \caption{\label{fig:QFI} Quantum Fisher information, $\mathcal{F}_{x}$, as a function of time for the RAP transfer from $\ket{S, S}$ to $\ket{S, 0}$ for three values of $\bar{\alpha}=\alpha/\chi^2$, $\bar{\alpha}_1 = 10^{-4}$, $\bar{\alpha}_2 = 10^{-2}$, and $\bar{\alpha}_3 = 10^{-1}$; $N=10$. We also present snapshots of the Wigner function of the collective state at the selected time.}
\end{figure}

To demonstrate the efficiency and robustness of the proposed scheme, in Fig.~\ref{fig:QFI}, we present the QFI time-evolution during the RAP process with several values of the chirp rate. Initially, the system is in the Dicke state $\ket{S, S}$, the QFI equals $N=10$. The QFI dynamics depend strongly on the chirp rate, yet all regimes give the same final result. For $\alpha=10^{-4}  \chi^2$, the dynamics of the Dicke state population is fully adiabatic; the population goes sequentially from one Dicke state to the next and, at most, only two adjacent states are populated at any time. In the plateau areas, the values of the QFI correspond to the QFI of individual Dicke states, $\mathcal{F}_{x} = N^2/2 - 2 m^2 + N$. The reduction of the QFI in this regime can be evaluated by calculating the QFI for a two-Dicke-state superposition $\ket{\psi} = \cos \zeta/2 \ket{S,m} +e^{i\phi} \sin \zeta/2 \ket{S, m\pm1} $. We find $\mathcal{F}_z =  \sin^2 \zeta$, $\mathcal{F}_{x} = N^2/2 + N - 2 m^2  + 2 (2 m \pm 1) \sin^2 \zeta/2 - \sin^2 \zeta \cos^2 \phi \left(N^2/4 - m^2 + N/2 \mp m\right)$, and $\mathcal{F}_{y} = N^2/2 + N - 2 m^2  + 2 (2 m \pm 1) \sin^2 \zeta/2 - \sin^2 \zeta \sin^2 \phi \left(N^2/4 - m^2 + N/2 \mp  m\right)$. These expressions explain the substantial reductions in the value of $\mathcal{F}_{x}$ when equal superpositions are created ($\zeta=\pi/2$), especially as $m$ decreases. The dynamics of the QFI in the $y$ direction, $\mathcal{F}_{y}$, (not shown here) is qualitatively similar and well correlated with the analytic expression above.

For larger chirp rates, $\alpha = 10^{-2} \chi^2$,  more Dicke states are populated, and the QFI reduces even further at intermediate times. For  $\alpha = 10^{-1} \chi^2$, we see that the QFI stays most of the time below the SQL ($\mathcal{F}_{x} = N$) since many Dicke states are populated simultaneously. However, a smooth switch-off of the coupling and the chirp guarantees the adiabatic passage to the target Dicke state.

A notable feature of the proposed RAP scheme is that the chirp rate $\alpha$ can be increased at least proportionally to $N$ (see Supplementary Material). Assuming a $\chi$ independent of $N$, we can conclude that the total time of the RAP scheme, which is roughly $N \tau = N \chi / \alpha$, can be independent of $N$. Moreover, in the limit where each transition is traversed at its maximum speed, the overall time of the RAP scheme could even decrease proportionally to $\log(N)/N$. The assumption that $\chi$ is independent of $N$ is possible for moderate values of $N$ (up to $N \sim 1000$) and could be accomplished by engineering the squeezing pulse, as discussed in \cite{ColomboNP2022,LiPRXQ2022} and the Supplementary Material. In the case of larger atom numbers, $\chi$ scales as $1/N$. Yet, the RAP is still efficient as long as the turn-on and -off time of the coupling field is adjusted to compensate for the reduction of the energy difference between adjacent Dicke states. In that case, the total time scales as $\log(N)$ for the non-negligible values of $\chi$. 

The above-described RAP scheme can be modified to prepare another class of correlated quantum states, providing sensitivity enhancement for Ramsey spectroscopic measurements. The metrological gain can be evaluated by the Wineland squeezing parameter~\cite{wineland1992spin, wineland1994squeezed}
\begin{equation}
  \label{eq:wineland}
  \xi^2
  = \Delta \varphi^2/\Delta \varphi_{\text{CSS}}^2
  =  \Delta \Op{S}_z^2 \, N/\Abs{\Avg{\Op{S}_x}}^2 \,,
\end{equation}
where $\Delta \varphi^2$ is the variance of a phase estimation for an entangled state and $\Delta \varphi^2_{\text{CSS}}$ is the result for a coherent state. Here, we have chosen $z$ as the squeezing direction and $x$ as the mean spin orientation.

From Eq.~\eqref{eq:wineland}, we can see that we need to minimize the quadrature in the $z$ direction, $\Delta \Op{S}_z$, while keeping the projection onto the $x$ axis, $\Avg{\Op{S}_x}$, as high as possible, since it defines the maximum contrast in the interferometric protocol. It has been shown~\cite{bloch2004control, toth2014quantum, sorensen2001entanglement, CarrascoPRA2022} that the optimal ESS states that minimize $\xi^2$ under the constraint of the fixed signal contrast must satisfy the equation  $[\chi \Op{S}_z^2 - \Omega \Op{S}_x] \ket{\Psi}_{\text{ESS}} = \lambda \ket{\Psi}_{\text{ESS}}$. Interestingly, as the signal contrast approaches zero, the ESS state becomes the Dicke state $\ket{S,0}$~\cite{toth2014quantum, CarrascoPRA2022}, and the squeezing parameter diverges. Indeed, the more metrologically useful the ESS states become, the more they approach the Dicke state $\ket{S,0}$, and are well-approximated by a linear combination of the Dicke states $\ket{S,0}$ and $\ket{S,\pm 1}$. For a fixed value of contrast, ESS states give HL scaling~\cite{CarrascoPRA2022}. Therefore,
creating them allows us to achieve such scaling for Ramsey interferometry.

During the RAP generating the $\ket{S=5, 0}$ Dicke state (Fig.~\ref{fig:to_m=0}), we observed transient population in the $\ket{S=5, \pm 1}$ states. To create the ESS state, we abruptly turn off the coupling, $\Omega (t)$, which results in some population of the $\ket{S, \pm 1}$ states at final time, thus creating the desired ESS state.

\begin{figure}
  \centering
  \includegraphics{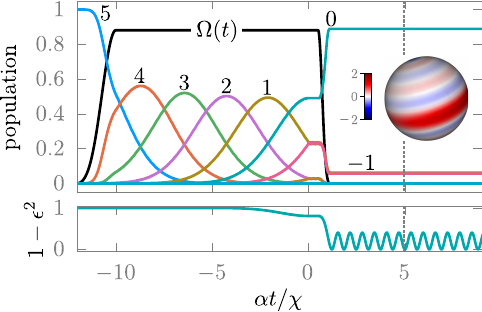}
  \caption{\label{fig:to_ESS} Generation of the ESS state via RAP, the chirp rate is $\alpha = 0.1\chi^2$. Upper panel: the Dicke state population dynamics, the time-dependent coupling with the maximum $\Omega_{\text{max}} = 0.88 \chi$ is shown by a black solid line. The Wigner function of the generated state at $t=5 \chi / \alpha$ is in the inset. Bottom panel: the infidelity as a function of time.}
\end{figure}

Figure~\ref{fig:to_ESS} shows the ESS state generation via fast turn-off of the coupling  during a RAP pulse aiming towards $\ket{S, 0}$. The main change in the time dependence of the coupling compared to the one in Fig.~\ref{fig:to_m=0} is that we set the switch-off time of the coupling pulse to $t_\text{off} = 0.583 \chi / \alpha$ and choose the turn-off time $t_2=0.5 \chi / \alpha$. The maximum overlap with the ESS target state is $\epsilon^2=0.9994$, while the averaged spin projection onto the $x$ axis is $\Avg{\Op{S}_x}=S/2$. Despite these modifications, a large parameter space region still gives excellent fidelity (see Supplementary Material).

Since the created ESS state is not an eigenstate of $\Op{S}_z^2$, which is the system's Hamiltonian after the coupling and the chirp are both turned off, the ESS state infidelity, $1- \epsilon^2$, oscillates with the frequency proportional to the shearing strength, $\chi$, as shown in the bottom panel of Fig.~\ref{fig:to_ESS}. The oscillations are relatively slow, and they end when turning off the OAT term in the Hamiltonian, Eq.~(\ref{Hamiltonian}), to achieve maximum fidelity. The ESS-state Wigner function is shown in the inset of Fig.~\ref{fig:to_ESS}. We observe a reduced variance in the $z$ direction, while there is a definite orientation of the total spin ($\Avg{\Op{S}_x}=S/2$) that ensures significant contrast of the Ramsey signal, as opposed to the case utilizing the Dicke state $\ket{S, 0}$.

\begin{figure}[tb]
  \centering
  \includegraphics{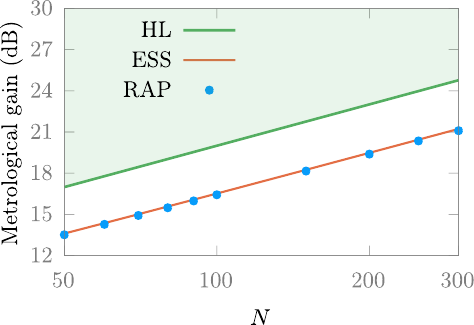}
  \caption{\label{fig:scaling} Comparison between the metrological gain as a function of the atom number for ideal ESS states and ESS states created by RAP. The results share the HL scaling.
  }
\end{figure}
So far, for illustration purposes, we have used only a small number of atoms. However, the proposed method works for arbitrary $N$. To demonstrate this, in Fig.~\ref{fig:scaling}, we plot the metrological gain of the RAP-produced state with respect to the CSS. As a target state, we use an ESS state with contrast $\Avg{\Op{S}_x} = S/2$. The metrological gain obtained with the RAP-produced states is practically identical to the ESS-state gain with the HL scaling.

To conclude, we have demonstrated the creation of many-atom entangled states via RAP between Dicke states. The generated $\ket{S,0}$-Dicke  and ESS states maximize the QFI and metrological gain for Ramsey interferometry. We have shown how to steer the system into the Dicke state $\ket{S, 0}$ and how to prepare an ESS state, providing HL scaling. The RAP technique is possible due to the unique structure of the non-linear OAT Hamiltonian. The process is exceptionally robust to driving field variations and variations in the number of atoms, eliminating the requirement of a precise count. In addition, the total time of the RAP is independent of $N$ for moderate $N$. These interesting properties open up the possibility of applying the RAP to create metrologically useful many-atom entangled states that are not easily accessible with other techniques, such as twist-and-turn strategies, that suffer from the accumulation of gates error~\cite{marciniak2022nature} and require substantial optimization efforts~\cite{CarrascoPRA2022} that become challenging  to implement as $N$ increases~\cite{ChiowPRL2011, KovachyN2015, Goerz2023Robust}. The technique could also work to prepare GHZ and various cat states. For instance, one could drive the system into $\ket{S,-S}$ (instead of $\ket{S,0}$) and adjust the turn-on of the pulses so that the first transition only transfers half of the population, thus creating a superposition of $\ket{S,S}$ and $\ket{S,-S}$. As an extension of this work, it could be beneficial to consider a shortcut-to-adiabaticity scheme~\cite{ChenPRL2010, Guéry-OdelinRMP2019} to speed up RAP, as well as applying advanced techniques of optimal quantum control~\cite{GoerzQ2022, SolaJPB2022} to maximize the fidelity (and metrological gain) and minimize losses due to decoherence, dephasing, and photon scattering. The remarkable robustness of RAP may also allow for the implementation of the protocol via Ising interactions that approximate the OAT Hamiltonian~\cite{franke2023quantum, eckner2023realizing, bornet2023scalable}, thus broadening the range of applications to other research areas. 

\begin{acknowledgments} 

This research was supported by DEVCOM Army Research Laboratory under Cooperative Agreement Numbers W911NF-21-2-0037 (SCC) and W911NF-17-2-0147 (MHG). VSM is grateful for support by a Laboratory University Collaboration Initiative (LUCI) grant from OUSD.

\end{acknowledgments} 

\bibliographystyle{apsrev4-2.bst}
\bibliography{refs}

%%%%%%%%%% Merge with supplemental materials %%%%%%%%%%
\widetext
\clearpage

% \cleardoublepage

% \newpage
% \appendix
% \onecolumngrid

% \pagebreak
% \widetext

\section{SUPPLEMENTARY MATERIAL}
\renewcommand{\theequation}{S\arabic{equation}}
\renewcommand{\thefigure}{S\arabic{figure}}
\setcounter{figure}{0}
\setcounter{equation}{0}

\subsection{\label{appendix:Robustness}Robustness analysis}

\begin{figure}[H]
  \centering
  \vspace{-5pt}
  \includegraphics{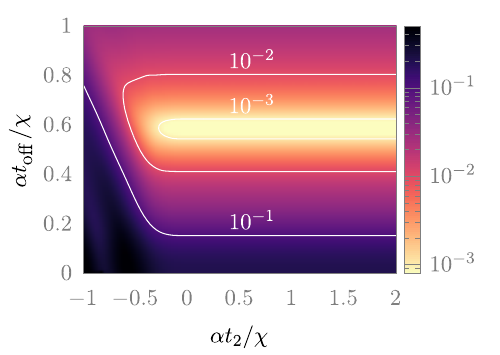}
  \caption{\label{fig:Robustness} Extreme spin squeezed state infidelity as a function of $t_2$ and $t_\text{off}$, the time where the coupling field plateau ends and the time that the coupling field takes to turn off completely, respectively. The calculations are performed for $N=10$ atoms, $\alpha = 0.1\chi^2$. and $\Omega_{\text{max}} = 0.88 \chi$.}
\end{figure}

In general, RAP schemes are well-known for being robust with respect to variations in the pulse parameters or imperfections in the pulse shape. In this section, we provide more details on the robustness of the proposed protocol. In Fig. \ref{fig:Robustness}, we show the infidelity in the creation of an extreme spin squeezed state as a function of $t_2$, the time where the coupling field plateau ends, and with $t_\text{off}$, the time that the coupling field takes to turn off completely. We observe a vast region of parameter space where the infidelity is reasonably small. Thus, upon choosing particular values of $t_2$ and $t_\text{off}$, a variation in $t_2$ and $t_\text{off}$ will result only in a slight variation in the fidelity.

\begin{figure}[b]
  \centering
  \vspace{-5pt}
  \includegraphics{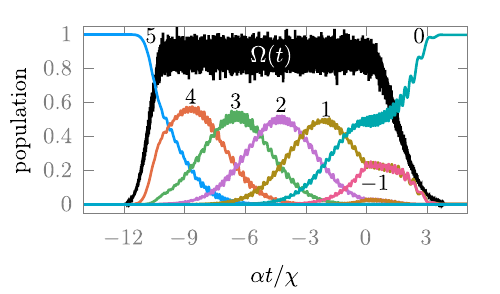}
  \caption{\label{fig:noisy_to_m=0} Population dynamics of the ten-atom Dicke states. The solid black line shows the coupling pulse shape, $\Omega (t)$, perturbed by multiplicative noise from a normal distribution centered in the unity with a standard deviation of $0.05$. The chirp rate is $\alpha = 0.1 \chi^2$.}
\end{figure}

Regarding imperfections in the pulse shape, in Fig. \ref{fig:noisy_to_m=0}, we show the population dynamics under our scheme for a noisy coupling field $\Omega (t)$. That noisy coupling is obtained by multiplying the original field with a noise factor drawn from a normal distribution centered at 1, with a standard deviation of $0.05$. At final time, a fidelity of $0.997$ is reached, demonstrating robustness to noise.

Let us also mention here that the effective (or normalized) "chirp" rate $\bar{\alpha}=\alpha/\chi^2$ can be varied by a few orders of magnitude without affecting the target state fidelity, as we demonstrated in Fig.~\ref{fig:QFI}. In summary, there is a wide range of parameter values such as coupling strength, ramp rate, and the time to turn on and off the coupling, where the adiabaticity condition of the RAP protocol are satisfied. Therefore, the robustness of the proposed scheme can facilitate future experimental implementations and testing.  

\subsection{Hamiltonian parameters and RAP scaling with the number of atoms \label{appendix:squeezing}}

To give an estimate of the proposed RAP scaling with the number of the atoms, we focus here on the Hamiltonian parameters $\chi$, $\beta(t)$, and $\Omega(t)$ scaling. The ramp function, $\beta(t)= \alpha t$, and the coupling, $\Omega (t)$, are, in a sense, single atom parameters defining the rate of the rotation over $z$ and $x$ axis and are independent of the number of atoms and can be controlled in an experiment. The value of the shearing parameter, $\chi$, is determined by the interactions between atoms and it is responsible for squeezing or creating entanglement in the atomic cloud. The shearing parameter value (the strength of the nonlinear term in the Hamiltonian) and scaling is defined by the experimental details of the one-axis twisting Hamiltonian engineering. Obviously, without the non-linear OAT term, the proposed RAP scheme cannot be implemented, and no squeezed or other entangled states can be prepared.  

\begin{figure}[tb]
  \centering
  \includegraphics{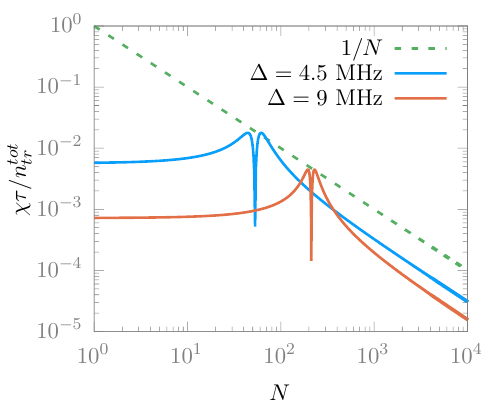}
  \caption{\label{fig:chi_example} Shearing parameter per transmitted photon rate. The solid lines show the dependence of $N$ for two values of detuning, according to Eq.~\eqref{eq:shearing}. For reference, we include a dashed line where we plot $1/N$, which is the scaling for $N \gg 1$.}
\end{figure}

According to Ref. \cite{ColomboNP2022}, where the implementation of an effective one-axis twisting via the nonlinear interaction between the atoms and light applied inside a cavity is discussed, the normalized shearing strength is given by
\begin{equation}
    \tilde Q = \sqrt{N} \chi \tau = \frac{n_{tr}^{tot}}{\sqrt{N}} \mathcal{L}_d(x_a) \mathcal{L}_a(x_a) \frac{\frac{N}{2} \eta^2 (1 + \frac{N}{2} \eta - x_a x_c)}{\left(1 + \frac{N}{2} \eta \mathcal{L}_a(x_a)\right)^2 + \left(x_c + \frac{N}{2} \eta \mathcal{L}_d(x_a)\right)^2} \ ,
\end{equation}
so, 
\begin{equation} \label{eq:shearing}
    \chi = \frac{n_{tr}^{tot}}{\tau N} \mathcal{L}_d(x_a) \mathcal{L}_a(x_a) \frac{\frac{N}{2} \eta^2 (1 + \frac{N}{2} \eta - x_a x_c)}{\left(1 + \frac{N}{2} \eta \mathcal{L}_a(x_a)\right)^2 + \left(x_c + \frac{N}{2} \eta \mathcal{L}_d(x_a)\right)^2} \  ,
\end{equation}
where $\eta$ is the single-atom cooperativity, $\mathcal{L}_a$ and $\mathcal{L}_d$ are Lorentzian profiles, $x_c=2\delta / \kappa$ is the detuning of the probe beam from the cavity resonance frequency normalized to the cavity linewidth, $x_a=2\Delta / \Gamma$ the detuning of the probe laser from the atomic resonance normalized to the atomic transition linewidth, $N$ is the number of atoms, and $n_{tr}^{tot}$ is the number of transmitted photons. In the limit $N \gg 1$, $\chi$ scales proportionally to $1/N$ for fixed values of the parameters. However, for low to intermediate values of $N$ (up to $N \sim 100$), $\chi$ remains fairly constant, as we can observe in Fig. \ref{fig:chi_example}, where we choose $\eta = 7.7$, the value measured in Ref. \cite{ColomboNP2022}, which could likely be improved to increase $\chi$. Indeed, the dependence with respect to $N$ can be attenuated by controlling the detuning as proposed in Ref. \cite{LiPRXQ2022} and experimentally demonstrated in Ref. \cite{ColomboNP2022}. For $\Delta = 9$ MHz, around the values explored in Ref. \cite{ColomboNP2022}, $\chi$ only decreases by a factor of $10$ for $N=1000$. Certainly, it is much better than the $N \gg 1$ limit suggests. Moreover, it could be compensated further by increasing the light intensity. Consequently, we can consider $\chi$ fairly independent of $N$ for low to moderate values of $N$. The upper limit of the atom number for which the shearing parameter can be considered constant can be adjusted in the experiment. 

\subsubsection{RAP scaling according to Landau-Zener model}

In the main text, we have provided the details of the RAP implementation and the dressed states picture in the $N$ atom system. To provide some additional insight into the RAP process and its time scaling with respect to the atom number, we now consider the simplified Landau-Zener model that can be applied to each crossing between the dressed Dicke states.

The probability of a diabatic transition, according to the Landau-Zener model, is given by
\begin{equation} \label{}
  P = e^{-2\pi \Gamma}
\end{equation}
where $\Gamma = a^2 / |\alpha|$, $a$ is the off-diagonal element of the Hamiltonian (coupling), and $\alpha = d(E_2-E_1)/dt$ is the rate at which the energies $E_{1,2}$ of the two crossing states (diabatic) change. That rate determines the speed of the overall process and is inversely proportional to the total time of the scheme.

Our case can be understood as a succession of adiabatic transitions. Indeed, for the transition between $\ket{N/2, m}$ and $\ket{N/2, m-1}$,
\begin{equation}
a = \frac{\Omega}{2} [(S+m) (S-m+1)]^{1/2} \,.
\end{equation}
Consequently,
\begin{equation}
  \Gamma_m = \frac{\Omega^2}{4 \alpha} (S+m) (S-m+1)\,,
\end{equation}
because
\begin{equation}
  \frac{d(E_m-E_{m-1})}{dt} = \frac{d(\alpha m t -\alpha (m-1) t)}{dt} = \alpha \,,
\end{equation}
as in the main text.

Now we consider the transition between $\ket{N_1/2, m_1}$ and $\ket{N_1/2, m_1-1}$, and the transition between $\ket{N_2/2, m_2}$ and $\ket{N_2/2, m_2-1}$. Those transitions happen with the same probability if $\Gamma_{m_1}(N_1) = \Gamma_{m_2}(N_2)$, therefore
\begin{equation}
  \frac{\alpha_2}{\alpha_1} = \frac{(N_2/2+m_2) (N_2/2-m_2+1)}{(N_1/2+m_1) (N_1/2-m_1+1)}\,.
\end{equation}
With $m_1=1$ and $N_1 = 2$, for reference, we obtain
\begin{equation} \label{eq:scaling}
  \frac{\alpha_2}{\alpha_1} = \frac{1}{2} (N_2/2+m_2) (N_2/2-m_2+1)\,.
\end{equation}
By considering $m_2 = 1$, we conclude that the transition between $\ket{N_2/2, 1}$ and $\ket{N_2/2, 0}$ can be sped up proportionally $N_2^2$ without losing fidelity. If $m_2 = N_2/2$, the transition speed-up is proportional to $N_2$.

\begin{figure}[tb]
  \centering
  \includegraphics[scale=0.97]{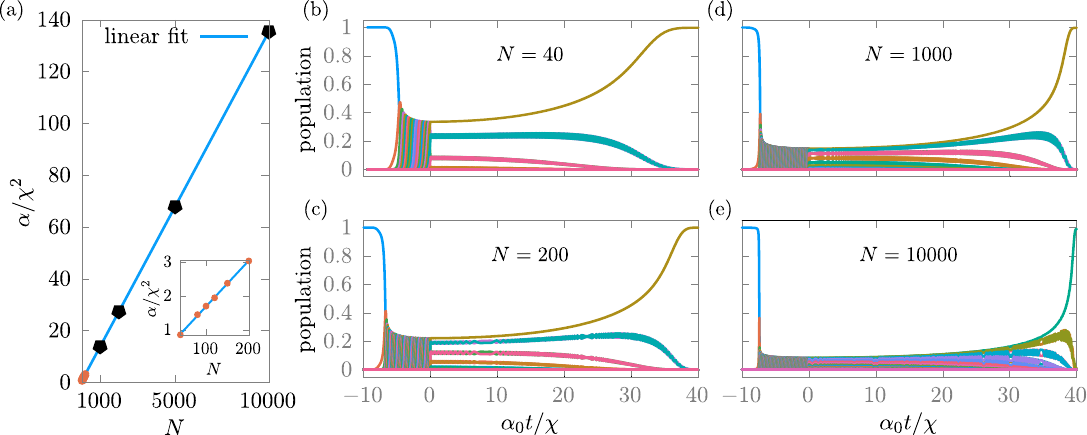}
  \caption{\label{fig:Alpha_scaling} Numerical results for the scaling of the chirp rate $\alpha$. (a) Maximum value of $\alpha$ that gives fidelity beyond $0.999$ as a function of $N$ for $\Omega_{\text{max}} = 0.88 \chi$. The circles are a result of numerical search for the value. The straight line correspond to a fit with a coefficient of determination of $0.9993$. The pentagons represent a simulation using the value of $\alpha$ extrapolated from the fit. In those simulations, we obtain fidelities beyond $0.99$. (b) Population dynamics for $N=40$ at the maximum chirp rate. (c) Population dynamics for $N=200$ at the maximum chirp rate. (d) Population dynamics for $N=1000$ at the chirp rate extrapolated from the fit. (e) Population dynamics for $N=10000$ at the chirp rate extrapolated from the fit. In panels (b), (c), (d) and (e), we use $\alpha_0 = 0.1\chi^2$ to scale time. Consequently, these four plots are in the same timescale.
  }
\end{figure}

In Fig. \ref{fig:Alpha_scaling}~(a), we corroborate Eq.~\eqref{eq:scaling} numerically by solving for the maximum value of $\alpha$ that gives fidelity beyond $0.999$ as a function of $N$ for $\Omega_{\text{max}} = 0.88 \chi$ (same parameters as Fig. 2 from the main text). We observe a linear dependence with $N$, as predicted. As discussed before, a superlinear dependency would break adiabaticity in the slowest transition. For reference, we include examples of population dynamics in panels (b) and (c). In panels (d) and (e), we include examples of the population dynamics for high values of $N$ where solving numerically for the maximum value of $\alpha$ is beyond our computational capabilities. Yet, the simulation using the chirp rate extrapolated from the linear fit reveals that the linear dependence still holds. We can observe that the population dynamics take roughly the same time despite the different values of $N$.

\subsubsection{Summary of the total time of the RAP scaling }

The proposed RAP protocol in the Dicke basis can be realized in several regimes. The first is a slow, sequential transfer from one Dicke state to the next, populating only two states during the process. That regime is mostly addressed in the first part of the paper, where we also explain the population transfer mechanism on the Dicke basis. In the proposed RAP scheme, the parameter $\alpha$ is the effective "chirp" rate, which defines the time of population transfer between the subsequent Dicke states according to the Landau-Zenner model. The time between sequential crossings is defined as $\tau=\chi/\alpha$. 

Therefore, in this regime (if we keep population transfer sequential, and assuming $\chi$ and $\alpha$ are independent of the number of atoms $N$), the total time of creating the Dicke state $m=0$ will scale linearly with a number of atoms and the crossing interval $\tau$. This regime might be impractical for a large number of atoms, but is still useful for understanding the Dicke state correlations and the RAP mechanism in systems with a moderate number of atoms. 

Making the RAP process much faster and keeping it adiabatic is also possible. In that fast regime, we do not restrict the number of Dicke states populated at any given moment of time, except at the beginning and the end of the process, where turn-on and turn-off stages should be properly adjusted. That regime can be realized by increasing the "chirp" rate $\alpha$ for the fixed value of $\chi$. Since for intermediate values of $N$, the shearing parameter $\chi$ can be considered independent of $N$, the total time of the RAP process would depend on $\alpha$. As discussed above, $\alpha$ can be increased proportionally to $N$ while retaining the fidelity of the scheme. Thus, we observe that the time between crossings decreases as $1/N$. That means that all crossings between Dicke states will take place almost simultaneously in the limit $N \gg 1$. In this case, the total time of the RAP process is independent of $N$ and is mostly defined by the turn-off and turn-on times of the coupling $\Omega(t)$. 
The above analysis also suggests that if each crossing is traversed at maximum speed, the RAP scheme time will decrease with $N$ proportionally to $\log(N)/N$, thus decreasing with $N$ at least for a moderate number of atoms. If $N$ is large enough to make $\chi$ scale as $1/N$, the RAP scheme time scales as $\log(N)$.
Fig.~\ref{fig:Alpha_scaling} shows the population dynamics of the Dicke states during the RAP process for $N=40$, $200$, $1000$, and $N=10000$ atoms. The adiabaticity criterion $\Omega^2_\text{max}\zeta_-^2 /\alpha \gg 1$ is perfectly satisfied in this regime.  Moreover, the adiabaticity condition is getting better for a larger number of atoms since the coupling parameter $\zeta_-^2$ in Eq.~(2) scales as $N$ or $N^2$ depending on the Dicke state transition. 

\begin{figure}[tb]
  \centering
  \includegraphics[scale=0.97]{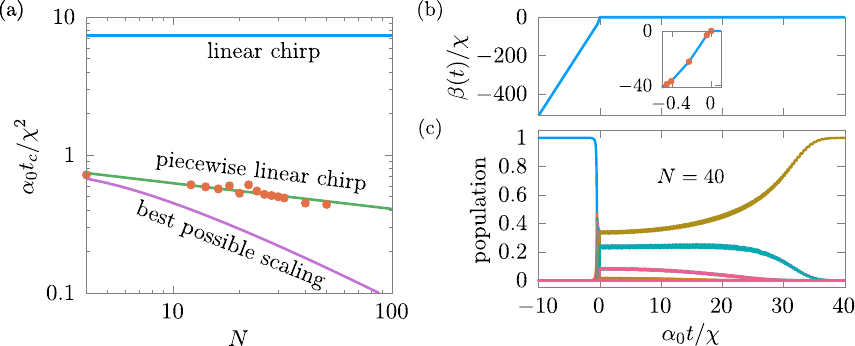}
  \caption{\label{fig:time_scaling}  RAP time scaling (time between the first and last crossing) for a linear and time-dependent chirp rate. (a) Comparison between the scaling for a linear chirp (constant), the best possible scaling (proportional to $\log(N)/N$), and proof-of-principle optimization of a variable chirp rate that gives fidelity beyond $0.999$ while minimizing the time between the first and last crossing (in points). The continuous line represents a power-law fit that shows scaling proportional to $N^{-0.186}$. (b) $\beta(t)$ for $N=40$. The inset shows a zoom to the optimized points that define the piecewise-linear ansatz (c) Population dynamics for $N=40$ for the optimized $\beta(t)$. In panels (b) and (c), we use $\alpha_0 = 0.1\chi^2$ to scale time. Consequently, these plots are in the same timescale of Fig.~\ref{fig:Alpha_scaling}.
  }
\end{figure}

Fig.~\ref{fig:time_scaling}~(a) graphically illustrates the scaling obtained using a linear chirp (no scaling with $N$), the best possible scaling ($\log(N)/N$), and the result of a proof-of-principle optimization of a variable chirp rate showing intermediate scaling proportional to $N^{-0.186}$. In this optimization, we assumed the chirp rate to be piecewise linear in the region of interest (where the crossings happen). We optimized the values of the breakpoints (see orange points in Fig.~\ref{fig:time_scaling}~(b) inset) to minimize the time between the first and the last crossing while retaining fidelity beyond $0.999$. The intermediate scaling obtained confirms that it is possible to traverse the crossing faster, as the theory predicts. More than that, the scaling changes, and the prefactor of such scaling improves an order of magnitude. Besides this, the numerical results also match the prediction from Landau-Zener theory that the first transition (from $m=S$ to $m=S-1$) is the slowest one, as the slope increases with time in Fig.~\ref{fig:time_scaling}~(b) (before stopping at $t=0$). Fig.~\ref{fig:time_scaling}~(c) shows an example of the dynamics for $N=40$ where we can observe that the whole process is much faster than its counterpart using a fixed chirp rate (see Fig.~\ref{fig:Alpha_scaling}~(b)).

The proposed RAP scheme clearly can be optimized further using methods of optimal control to improve scaling, especially the on- and off-stages of the coupling. Additionally, the RAP results can be considered using a variety of methods that have been developed under the name of ``shortcuts to adiabaticity'', to minimize the total time of the process.     

\subsection{Evaluation of the Wigner function \label{appendix:wigner}}
The Wigner function for a system of spin $S$ is given by
\begin{equation}
    W(\theta, \phi) = \sqrt{\frac{2\pi}{S}} \sum_{\ell=0}^{2S} \sum_{m=-\ell}^{\ell} G_{\ell, m} Y_{\ell, m} (\theta, \phi) \,
\end{equation}
where $Y_{\ell, m}$ are spherical harmonics functions. The coefficients of this expansion are obtained from the orthogonality of the multipole operators $G_{\ell, m} = \text{Tr}[\rho \mathcal{T}_{\ell, m}^\dagger]$ where $\rho$ is the density matrix,
\begin{equation}
    [\mathcal{T}_{\ell, m}]_{m_1, m_2} = \sqrt{(2\ell + 1)/(2S+1)} \, C^{S, m_1}_{S, m_2, \ell,m} \, ,
\end{equation}
and $C^{S, m_1}_{S, m_2, \ell, -m}$ are the Clebsch-Gordan coefficients. Efficient numerical implementations to evaluate the Wigner function can be found in~[55].

\end{document}